\documentclass[sigconf]{acmart}

\usepackage{amsmath} 
\usepackage{bbold}
\usepackage{color}
\usepackage{subcaption}
\usepackage{multirow}
\AtBeginDocument{%
  }

\setcopyright{acmlicensed}
\copyrightyear{2024}
\acmYear{2024}
\acmDOI{XXXXXXX.XXXXXXX}

\begin{document}

\title{Personalised Outfit Recommendation via History-aware Transformers}

\author{Myong Chol Jung}
\affiliation{
\institution{Amazon Machine Learning}
  		\city{Melbourne}
		\country{Australia}
}
\authornote{Work done while at Amazon.}
\email{david.jung@monash.edu}

\author{Julien Monteil}
\affiliation{%
\institution{Amazon Machine Learning}
  		\city{Brisbane}
		\country{Australia}
}
\email{jul@amazon.com}

\author{Philip Schulz}
\affiliation{%
\institution{Amazon Machine Learning}
  		\city{Sydney}
		\country{Australia}
}
\email{phschulz@amazon.com}

\author{Volodymyr Vaskovych}
\affiliation{
\institution{Amazon Machine Learning}
  		\city{Melbourne}
		\country{Australia}
}
\email{vaskovyc@amazon.com}

\renewcommand{\shortauthors}{Jung et al.}

\begin{abstract}
We present the history-aware transformer (HAT), a transformer-based model that uses shoppers' purchase history to personalise outfit predictions. The aim of this work is to recommend outfits that are internally coherent while matching an individual shopper's style and taste. To achieve this, we
stack two transformer models, one that produces outfit representations
and another one that processes the history of purchased outfits for a given shopper. We use
these models to score an outfit's compatibility in the context of 
a shopper's preferences as inferred from their previous purchases.
During training, the model learns to discriminate between purchased and random outfits using 3 losses: the focal loss for outfit compatibility typically used in the literature, a contrastive loss to bring closer learned outfit embeddings from a shopper's history, and an adaptive margin loss to facilitate learning from weak negatives. Together, these losses enable the
model to make personalised recommendations based on a shopper's purchase history.

Our experiments on the IQON3000 and Polyvore datasets show that HAT
outperforms strong baselines on the outfit Compatibility Prediction (CP)  and the Fill In The Blank (FITB) tasks. The model improves AUC for the CP hard task by 15.7\% (IQON3000) and 19.4\% (Polyvore) compared to previous SOTA results.  It further improves accuracy on the FITB hard task by 6.5\% and 9.7\%, respectively. We provide ablation studies on the personalisation, constrastive loss, and adaptive margin loss that highlight the importance of these modelling choices.
\end{abstract}



\received{10 August 2024}
\received[revised]{12 March 2009}
\received[accepted]{5 June 2009}

\maketitle

\section{Introduction}

Fashion shoppers want stylistically matching outfits that look good. In a brick-and-mortar store, a sales assistant can help coordinating an outfit. In an online setting, shoppers rely on outfit recommendations provided on a website. To showcase relevant outfits from millions of individual fashion items, e-commerce retailers need to automate their recommendations. Consequently, outfit recommendation is a task that has been approached in several ways in the research literature, e.g., \cite{sha2016approach, sarkar2022outfittransformer, li2017mining, lorbert2021scalable, zheng2021collocation,li2020learning}. However, these recommendations are generic and 
reflect outfit preferences aggregated across shoppers. To match the 
physical store experience, shoppers need personalised outfit recommendations
that reflect their individual taste and style. 


Existing studies like~\cite{sarkar2022outfittransformer,li2020learning} have addressed non-personliased outfit recommendation problems by optimising two distinct tasks which are: (1) the outfit Compatibility Prediction (CP) task that predicts whether items in an outfit are compatible or not, and (2) the Fill in the Blank (FITB) task that selects the most compatible item for an incomplete outfit given a set of candidate choices. However, these approaches do not relate the outfit to the preferences of the shopper who is looking at it. This can lead to recommendations that fit well for shoppers with mainstream taste but are sub-optimal for all other customers. 

To overcome this limitation, we propose the history-aware transformer (HAT), a personalised stacked transformer model that incorporates a shopper's view or purchase history. In order to enable personalisation, we introduce a contrastive loss to bring the embeddings of outfits that have been bought by the same shopper close to each other. We further introduce weak negative outfits which are purchased or curated outfits in which single item has been randomly replaced. These weak negatives make the discrimination task harder, meaning that bought-together outfits form tighter clusters in latent space. This form of data augmentation is inspired by the intuition that most shoppers are looking to buy an item that is compatible with their existing selection instead of buying a whole outfit. 

Finally, we introduce an adaptive margin ranking loss that forces the compatibility score of a weak negative outfit to be lower than the compatibility score of a corresponding positive outfit by a certain margin. This margin should be larger if a stylistically important item is replaced in the weak negative and smaller if a less important item is switched. The importance of each item in an outfit is computed by a cross-attention mechanism with learnable queries, inspired by the Q-former~\cite{li2023blip}. The full HAT model is trained with positive, negative and weak negative labels (ratio 1:1:1) on a weighted combination of the focal loss, contrastive loss, and adaptive margin ranking loss.

In summary, our contributions are four-fold: 
\begin{enumerate}
    \item We propose HAT, a stacked transformer architecture that jointly learns individual outfit embeddings while also summarising a shopper's purchase history.
    \item We enable personalisation based on purchase history by utilising a contrastive loss to outfit embeddings bought by the same shopper closer to each other.
    \item We explicitly model the fact that shoppers desire to buy items that are compatible with their existing selection, by introducing weak negatives, that consists of bought outfits with 1 item randomly replaced, and an adaptive margin ranking loss to adapt to the importance of the replaced items.
    \item We show that our approach outperforms competitive baselines on the  IQON3000 and Polyvore datasets by 15.7\% and 19.4\% in terms of AUC on the outfit Compatibility Prediction (CP) hard task, and by 6.5\% and 9.7\% in terms of accuracy on the Fill In The Blank (FITB) hard task.
\end{enumerate}

\section{Related Work}
\subsection{Outfit Recommendation}
\subsubsection{Non-personalised outfits}
Fashion recommenders is a subset of recommendation systems that involve a particular domain market: garments and fashion items. 
These systems typically face problems such as data sparsity caused by the large amount of unique items or a lack of detailed product information. These problems are further aggravated when recommending outfits which are combinations of coherent individual fashion items. This makes traditional recommendation systems \cite{yehuda2009matrix, Schafer2007collaborative} based on user-item interactions challenging to use. 

The most straightforward approach to obtain outfit-level representations is to use multi-instance pooling on individual item embeddings \cite{li2017mining, tangseng2017recommending}. While this is an easy and efficient technique, it cannot capture complex outfit-item interactions. To fix this shortcoming, sequence, graph, and attention models have been proposed. In sequence modelling \cite{han2017learning, dong2019personalized, jiang2018outfit, li2017mining, lorbert2021scalable}, outfits are presented as an ordered sequence of items and variations of RNNs are used to obtain a final outfit representation. However, this approach makes strong assumptions on the item order while outfits are inherently unordered as conceded in~\cite{lorbert2021scalable}. 

As an improvement to sequence modelling, 
attention-based approaches have gained popularity in the recent years. Content Attentive Neural Networks (CANN) were proposed in~\cite{li2020learning}, where attention blocks are used to find representations responsible for compositional coherence between global contents of outfit images and coherence from semantic-focal contents. 
Mixed Category Attention Net (MCAN) was introduced in~\cite{yang2020learning} which leverages item category information in its attention networks to find better recommendation. OutfitTransformer \cite{sarkar2022outfittransformer} feeds an item's text and image embeddings into the transformer block together with a learnable outfit token. The output of the transformer encoder serves as the global outfit representation. The outfit representation is fed into an MLP used to judge the  compatibility of the outfit. While the OutfitTransformer does not make 
personalised predictions, we leverage part of this architecture for the personalised outfit tasks (see Section~\ref{sec:HAT}).

\subsubsection{Personalised outfits}
A notable drawback of non-personalised outfit recommendation systems is that the shopper's individual preferences are not considered when  recommendations are made. Thus, many studies have proposed additional components to the outfit recommendation systems to incorporate personal preference signals.

Similar to non-personalised recommendation, there were attempts \cite{Lu2019Learning, dong2019personalized, Lu2021personalized, zhan2021pan} to use pairwise item-item compatibility together with shopper-item compatibility. 
~\citet{lin2020outfitnet} proposed OutfitNet, a two-stage approach that first learns item compatibility of embeddings which then applies attention-based pooling to shopper and item embeddings to compute outfit relevancy scores. The model uses a triplet loss to maximise the difference between the relevancy scores of outfits that shoppers bought and ignored.

Similarly, 
Personalised Outfit Generation (POG) was proposed in \cite{chen2019pog} for tackling the outfit generation problem. They use transformer blocks on embeddings of all items in the outfit and a masked embedding for the missing item. The output embedding on the masked slot is used to find the best matching item out of the alternatives. Contrary to our work, personalised outfit generations is achieved by encoding the entire purchase history of a shopper as one outfit with no masks. This embedding is then used in the masked transformer decoder blocks to generate outfit items, one at a time. 

\begin{figure}[t]
\centering
\includegraphics[width=0.47\textwidth]{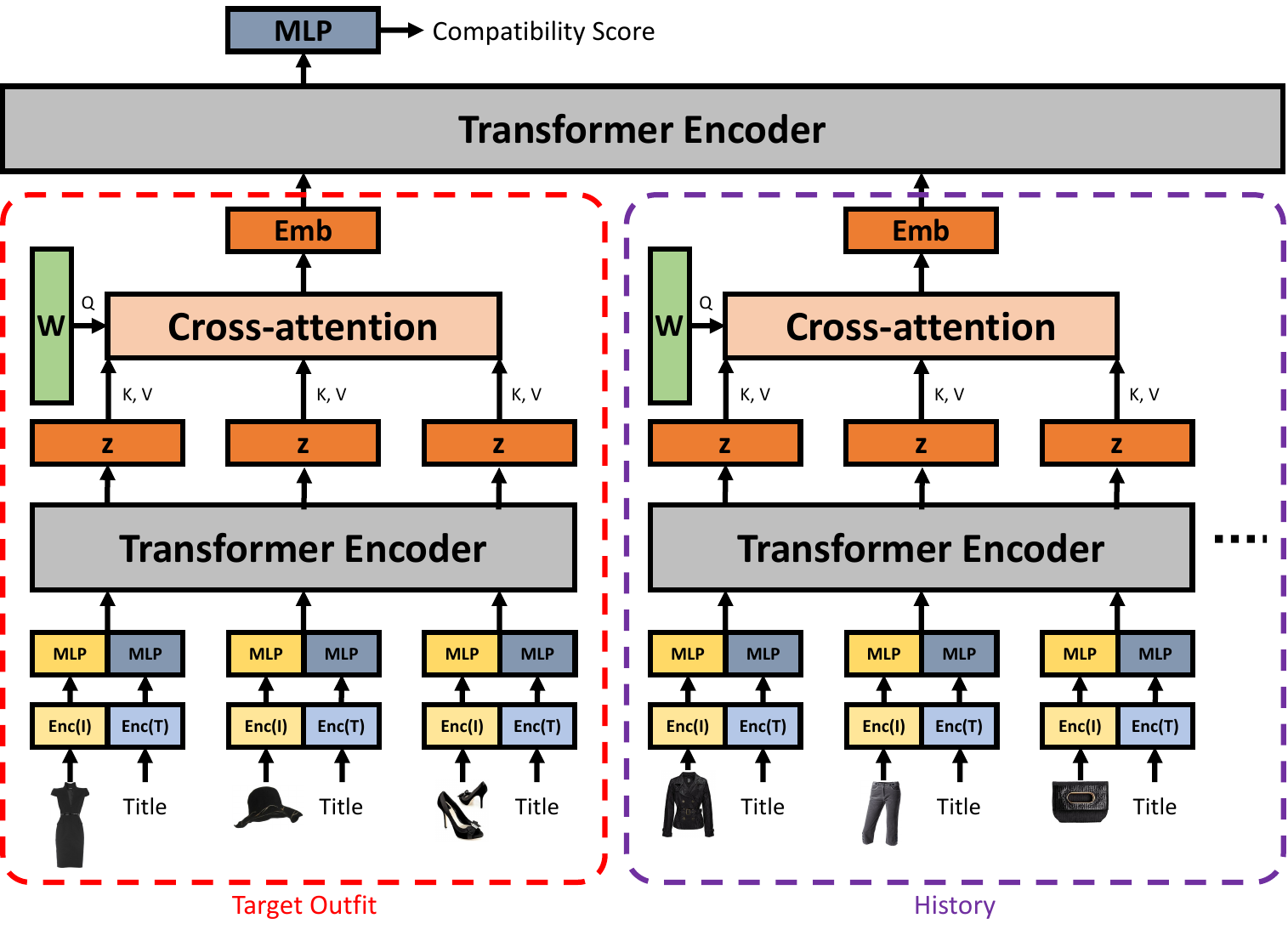}
\caption{Model diagram of HAT. The target outfit is an outfit whose compatibility score we wish to compute. It consists of images and titles for each item. K, V, and Q represent key, value, and query respectively for the cross-attention. The bottom-level transformer encoder shares weights.}
\Description{Model diagram.}
\label{fig:model diagram}
\end{figure}

\subsection{Sequential Recommendation}

While the focus of this paper is personalised outfit recommendation, the
personalisation mechanism can be viewed as a form of sequential 
recommendation, an active field of recommendation research. Before reviewing the relvevant literature, let us emphasise that all works known to us focus on next-item prediction and do not address the problem
of recommending outfits or, more generally, sets of products. 

Many sequential approaches are based on Wide and Deep \cite{wideanddeep} in that they
process the history of user-product interactions in separate sub-network
whose output is later combined with output based on other features. 
Prominent sequential recommenders include Deep Interest Networks 
\cite{DIN:2018} and their successor, the Behavior Sequence Transformer
\cite{behaviorTransformer:2019}. In general, there has been vivid
interest in using transformer models for sequential recommendation 
\cite{bert4rec, sasrec}. The most similar work to our approach is SSE-PT \cite{sse-pt} which uses a transformer to encode a user history and predict the last item in that history. Contrary to SSE-PT, our work focuses on scoring the consistency of outfits, does not use non-standard regulariasation schemes and does not rely on user embeddings, making it
more likely to perform well on cold customers.

Finally, using pre-
trained LLMs for sequential recommendation is alluring but has proven to be challenging \cite{liu2023chatgpt}. However,
recent work that combines purpose-trained transformers with LLMs is 
showing the feasibility of combining recommendation knowledge with the 
strong reasoning abilities of LLMs \cite{userllm, qu2024elephant}. It remains to be seen if the ideas presented in those works carry over to 
set/outfit recommendation.

\subsection{Contrastive Learning}

Constrastive learning was first used in \cite{Gutmann2010} to estimate intractable distributions by turning the density estimation problem into a classification problem. Since this seminal work, contrastive learning has been further developed by e.g.~\cite{InfoNCE} to extend to a setting where each positive sample is paired with multiple negative samples. This makes the classification problem harder and thus leads to better representation learning, and has since been successfully applied in several representation learning frameworks such as CLIP~\cite{clip:2021}. Supervised contrastive learning~\cite{supervised-contrastive} allows for more than one positive sample by using class labels. Each item in a batch serves as an anchor; all other items with the same class label are treated as positive while all remaining samples are treated as negative. Supervised constrastive learning seeks to maximise the inner product between the vector representation of the anchor and all positive samples. It has the effect that normalised embeddings from the same class are pulled closer together than embeddings from different classes. In practice, negative samples are typically drawn from the same mini-batch instead of the entire dataset. 


\section{Methodology}

In this section, we present the details of the proposed architecture for personalised outfit recommendation and the corresponding training procedure. We also present a data augmentation strategy that uses ``weak negative'' outfits in training (Sec~\ref{sec:adaptive margin loss}). Following previous work \cite{sarkar2023outfittransformer}, we represent an outfit through the titles and 
images of the individual items it contains. To add personalisation to our
model, we represent a shopper's style through the outfits they have purchased in the past. Technically, we aim to predict the
style consistency of the target outfit conditioned on the outfits in a shopper's interaction history.

\subsection{Problem Formulation}
We define the $o^{th}$ outfit in a shopper's history as $S_o=\{x^o_i\}_{i=1}^{N^o}=\{I^o_i, T^o_i\}_{i=1}^{N^o}$ where $x^o_i$ is an individual fashion item represented by an image $I^o_i$ and a title $T^o_i$ with $N^o$ being the number of items in the outfit. We denote the $u^{th}$ shopper's purchase history by $H_u=\{S_o\}_{o=1}^{M^u}$ where $M^u$ is the number of outfits purchased by the shopper. Our task is to predict a compatibility score (of whether the items in an outfit match together) $p^o\in [0,1]$ for a target outfit $ S_t $. During training, the ground truth for the target is either $y^o=1$ for an outfit judged as compatible (e.g. via annotation) or $y^o=0$ for an incompatible outfit (e.g. a random  outfit). The score $ p^0 $ represents the compatibility of an outfit as 
predicted by our model.

The different layers in our model architecture are denoted by lower case letters. We use greek subscripts to denote model parameters. For example, $ e_\theta $ and $ e_\phi $ are the image and title encoders.

\subsection{History-aware Transformer (HAT)}
\label{sec:HAT}

We propose the History-aware Transformer (HAT) model to personalise outfit compatibility scoring, shown in Fig.~\ref{fig:model diagram}. HAT is a two-level stack of transformer encoders~\cite{vaswani2017attention}. The bottom-level transformer encoder generates an outfit embedding for the purpose
of understanding the internal compatibility of that outfit. The top-level transformer encoder enables personalisation. It computes the compatibility score of the target outfit in the context of a shoppers's previously purchased outfits. By incorporating information about the purchased outfits as inputs to the top-level encoder, the model can learn not only whether a given outfit is compatible on its own but also if the given outfit matches well with the other outfits that the shopper has purchased. This allows the model to match the current outfit to the shopper's fashion style as inferred from their historical purchases.

\subsubsection{Bottom-level Encoder} 
For a given outfit $ S_o $, the bottom-level transformer encoder $ t_{\psi} $ processes item embeddings which are concatenations of an encoded image $e_{\theta}(I^o_i)$ and an encoded title $e_{\phi}(T^o_i)$. In practice,
we use pre-trained CLIP models \cite{clip:2021} to embed both the images and titles.
The outputs of the bottom-level transformer encoder, representing each item as $z^o_i$, are fed into a cross-attention layer. We introduce a learnable weight ($W$) as a query which is shared between all outfits. This query lets the model attend to important items in an outfit. Since $ W $ is learned, we make no prior assumptions about what item types (e.g. trousers vs. shirts) are most important for the compatibility of an outfit. We call  the resulting outfit embedding $E^o$. Formally, we define the outfit embedding of the $o^{th}$ outfit as:
\begin{gather}
    E^o = A^o Z^o = \text{Softmax}\left(\frac{W{Z^o}^{T}}{d}\right)Z^o \label{eq:outfit embedding} \\ 
    W \in \mathbb{R}^{d}, \quad Z^o  \in \mathbb{R}^{N^o\times d} \nonumber
    \\ 
     Z^o=\left[z^o_1, \ldots, z^o_{N^o} \right]=t_{\psi}\left(\left[e_{\theta}(I^o_i);e_{\phi}(T^o_i)]\right]_{i=1}^{N^o_{item}}\right)
\end{gather}
where $A^o$ is the cross-attention matrix, $d$ is the latent dimension, $t_{\psi}(\cdot)$ is the bottom-level transformer encoder with learnable parameters $ \psi $, and $[\cdot;\cdot]$ is concatenation along the feature dimension. The bottom-level transformer encoder $ t_{\psi}$ captures the relationship between the item embedding within an outfit. It is largely
inspired by the OutfitTransformer \cite{sarkar2023outfittransformer}; the
addition of the attention mechanism to weigh an item's importance within the outfit via a learnable query is novel. As we show in Tables~\ref{tab:CP performance} and~\ref{tab:num history ablation}, it improves upon the original non-personalised Outfit Transformer.

Conceptually, there are two advantages of introducing the learnable outfit representation using cross-attention. Firstly, the resulting outfit representation is a single vector independently of the number and the order of items. Secondly, the attention matrix approximates the importance of each item in an outfit (e.g., a top may have higher contribution to the compatibility of an outfit than an earring) without making prior assumptions. We show how to use the 
attention weights in the adaptive margin loss we introduce in Section~\ref{sec:adaptive margin loss}.

\subsubsection{Top-level Encoder} 
The inputs of the top-level transfomer encoder $ t_{\rho}$ are outfit representations $ \{E^i\}_{i=1}^{M_u} $ from a user's purchase history $ H_u $ generated by the bottom-level encoder $ t_\psi$. In particular, we concatenate a representation of the target outfit (i.e. the outfit whose compatibility we want to estimate) and the representations from outfits in a shopper's the purchase history. We then use an MLP 
$ f_\lambda $ to predict the compatibility score. We use the sigmoid function $ \sigma $ to bring the prediction onto the unit scale.

\begin{equation}
    p^o = \sigma\left(f_\lambda\left(t_\rho\left([E_t; E_1, \ldots, E_{M_u}]\right)\right)\right)
\end{equation}

By leveraging the shopper's purchase history, the model is able to estimate the compatibility of the target outfit $ S_t $ (whose outfit represenation is $ E_t $) in the context of the shoppers's personal style. Notice that the target outfit is not included in the history; it can be varied to score different target candidates.

\subsection{Model Training}

In this section, we discuss how to train HAT for personalised
outfit recommendation. The training procedure contains three losses which
address different aspects of the personalised recommendation problem. We define all losses over a mini-batch $ B $ of size $|B| = K$.

\subsubsection{Outfit Compatibility}
We follow \cite{sarkar2023outfittransformer} in using the focal loss~\cite{lin2017focal} for compatibility prediction:
\begin{align}
    \mathcal{L}_{FL}\left(B\right)&= \sum_{k \in B} F(p^o) \text{;  where} \nonumber \\
        F &= \begin{cases}
            -\alpha(1-p^o)^\gamma \log{(p^o)},& y^o=1 \\
            -(1-\alpha)({p^o})^\gamma \log{(1-p^o)},& y^o=0
        \end{cases}
\end{align}
where $\alpha$ and $\gamma$ are hyperparameters that balance classes and difficult samples respectively.

We limit the maximum number of history outfits since long histories are rare but take up a lot of GPU memory. If the total number of outfits purchased by a shopper exceeds the threshold, we randomly sample outfits. Refer to Section~\ref{sec:number of history} for the impact of the number of history outfits on performance.

\subsubsection{Personalisation by Contrastive Learning}

\begin{figure}
\centering
\includegraphics[width=0.30\textwidth]{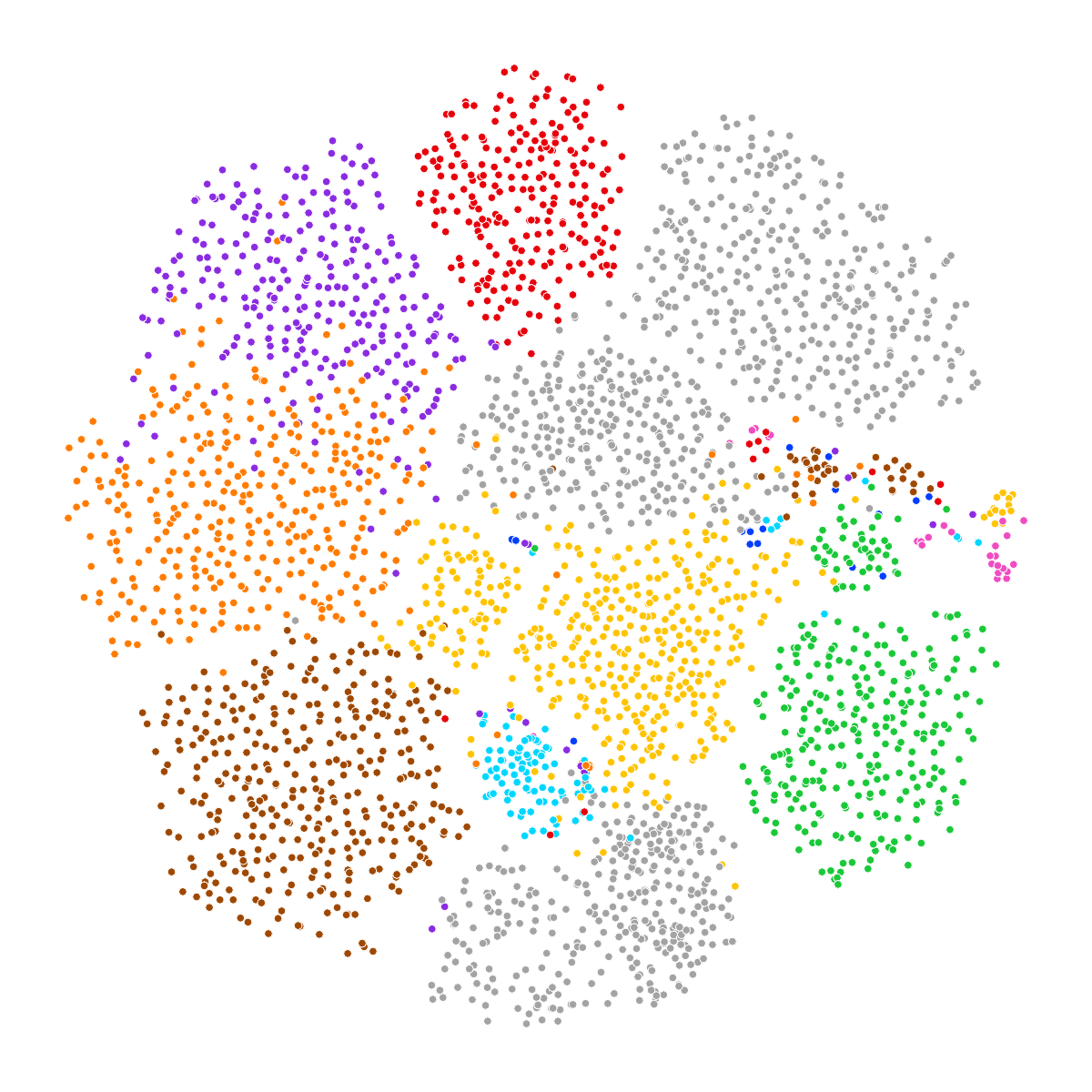}
\caption{Exemplary outfit embeddings projected to 2D space by t-SNE~\cite{van2008visualizing} in IQON3000. Colours indicate different shoppers.}
\Description{Exemplary outfit embeddings projected with t-SNE~\cite{van2008visualizing}.}
\label{fig:outfit embedding tsne}
\end{figure}

In order to fully utilise the top-level transformer, we found that it is important to bring the embeddings of outfits that have been bought by the same shopper close to each other. This produces outfit representations $E^o$ for which the compatibility score is dependent on the outfits in the history. To achieve this, we leverage supervised contrastive learning~\cite{khosla2020supervised}. Each history outfit embedding in a mini-batch is an anchor, outfit embeddings from the same shopper are positives, and outfit embeddings from different shoppers are negatives. In other words, we leverage shoppers as labels for supervised contrastive learning.

Let $ E^k $ and $ E^p $ be outfit representations within batch $ B $. During training, a mini-batch contains purchase histories of various shoppers. Let $ID(k)$ be the shopper ID associated with the $k^{th}$ outfit. We define the positive samples for the $ k^{th} $ outfit in the batch as $P(k)\equiv \{p\in B^{-k}: ID(p)=ID(k)\}$, i.e. those outfits from the same shopper's purchase history. We further define $ B^{-k} = B \setminus \{k\} $ to be the batch without the $ k^{th} $ outfit. The supervised contrastive loss is defined as:
\begin{equation}
    \mathcal{L}_{CL}\left(B\right)=\sum_{k\in B}\frac{-1}{\vert P(k) \vert} \sum_{p \in P(k)} \log\left({\frac{\exp{(E^k \cdot E^p/\tau)}}{\sum_{n \in B^{-k}}\exp{(E^k \cdot E^n/\tau)}}}\right)
\end{equation}
where $\tau$ is a temperature scale parameter. The effect of the supervised contrastive loss is that outfit representation from the same shopper's history are close to each other. This is illustrated in Fig.~\ref{fig:outfit embedding tsne}. Refer to Section~\ref{sec:Contrastive learning} for an ablation study on the effectiveness of the supervised contrastive loss.

\subsubsection{Adaptive Margin Loss}
\label{sec:adaptive margin loss}
\begin{figure}
\centering
    \begin{subfigure}{0.3\textwidth}
        \centering
        \includegraphics[width=\textwidth]{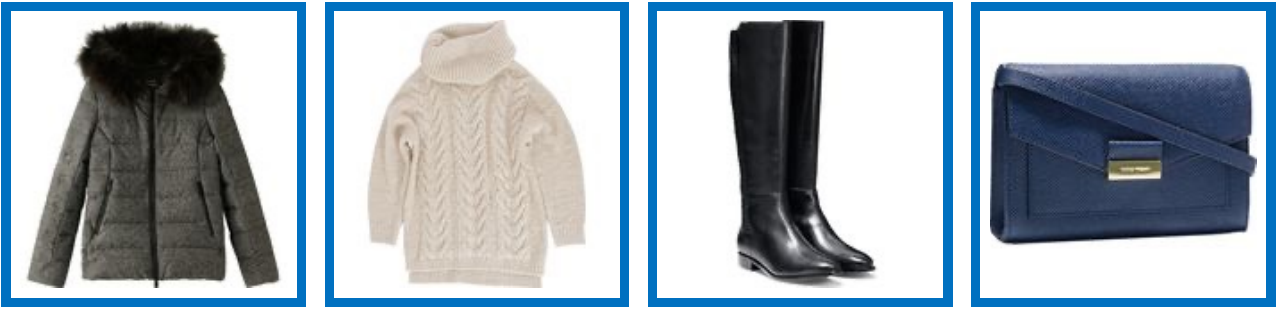}
        \caption{Positive outfit.}
        \label{fig:positive sample}
    \end{subfigure}
    \begin{subfigure}{0.3\textwidth}
        \centering
        \includegraphics[width=\textwidth]{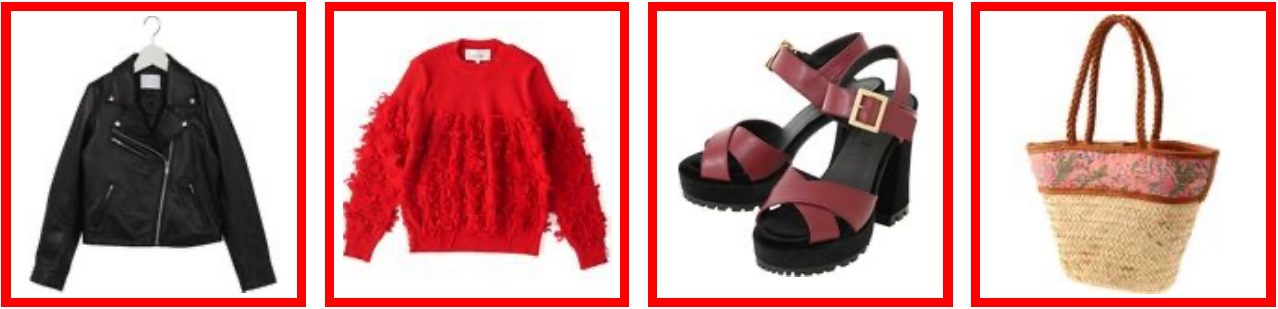}
        \caption{Negative outfit.}
        \label{fig:negative sample}
    \end{subfigure}
    \begin{subfigure}{0.3\textwidth}
        \centering
        \includegraphics[width=\textwidth]{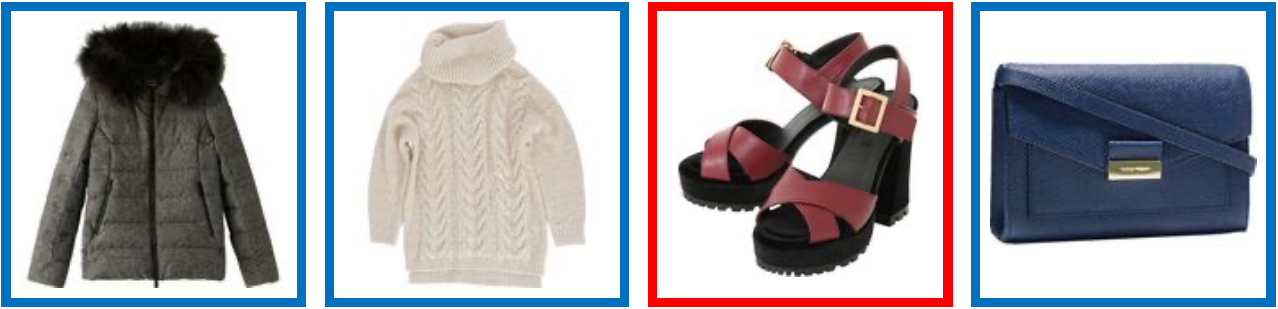}
        \caption{Weak negative outfit.}
        \label{fig:weak negative sample}
    \end{subfigure}
\caption{Examples of positive outfit (a), negative outfit (b), and weak negative outfit (c). Every item of the negative outfit is randomly selected item in place of the positive items within the same item category. On the other hand, a weak negative outfit only has a single item switched.}
\label{fig:samples}
\end{figure}

Previous studies \cite{sarkar2022outfittransformer} have created negative outfits by randomly switching all items of an outfit as shown in Fig.~\ref{fig:positive sample} and Fig.~\ref{fig:negative sample} to train outfit recommendation models. While this helps the model distinguish an annotated outfit and a random outfit, the signal is weak since random assortments of items are likely to be non-compatible and hence easy to distinguish from curated outfits. A more challenging task is to distinguish the positive outfit and a ``weak negative'' sample in which only one of items is randomly switched as shown in Fig.~\ref{fig:weak negative sample}. This is also closer to the choice that shoppers face when buying clothes. Most of the time, they are looking to buy an item that is compatible with their existing wardrobe; a shopper may look for a pair of shoes that match well to a top and a bottom that he already owns. Thus, we augment our training data with weak negative outfits while training the model to learn fine-grained details of a compatible outfit.
 
When using weak negative outfits for model training, one challenge is the lack of ground truth compatibility for these outfits. It would be misleading to label them as either 0 (fully incompatible) or 1 (fully compatible). To address this issue, we propose a margin ranking loss which requires the compatibility score of a weak negative sample to be lower than compatibility score of a corresponding positive sample by a certain margin. This margin ranking loss is defined as:
\begin{equation}
    \mathcal{L}_{M}\left(B\right) = \sum_{k \in B}(0,-p_p^k+p_w^k+m)
    \label{eq:margin loss}
\end{equation}
where ($x, y$) denotes a maximum value between $x$ and $y$, $p_p^k$ is the compatibility score for a positive (original) version of outfit $ o^k $ and $p_w^k$ is the compatibility score for a weak negative version of outfit $ o_i $. We use $m$ to denote the fixed margin. 

\begin{figure}
\centering
\includegraphics[width=0.43\textwidth]{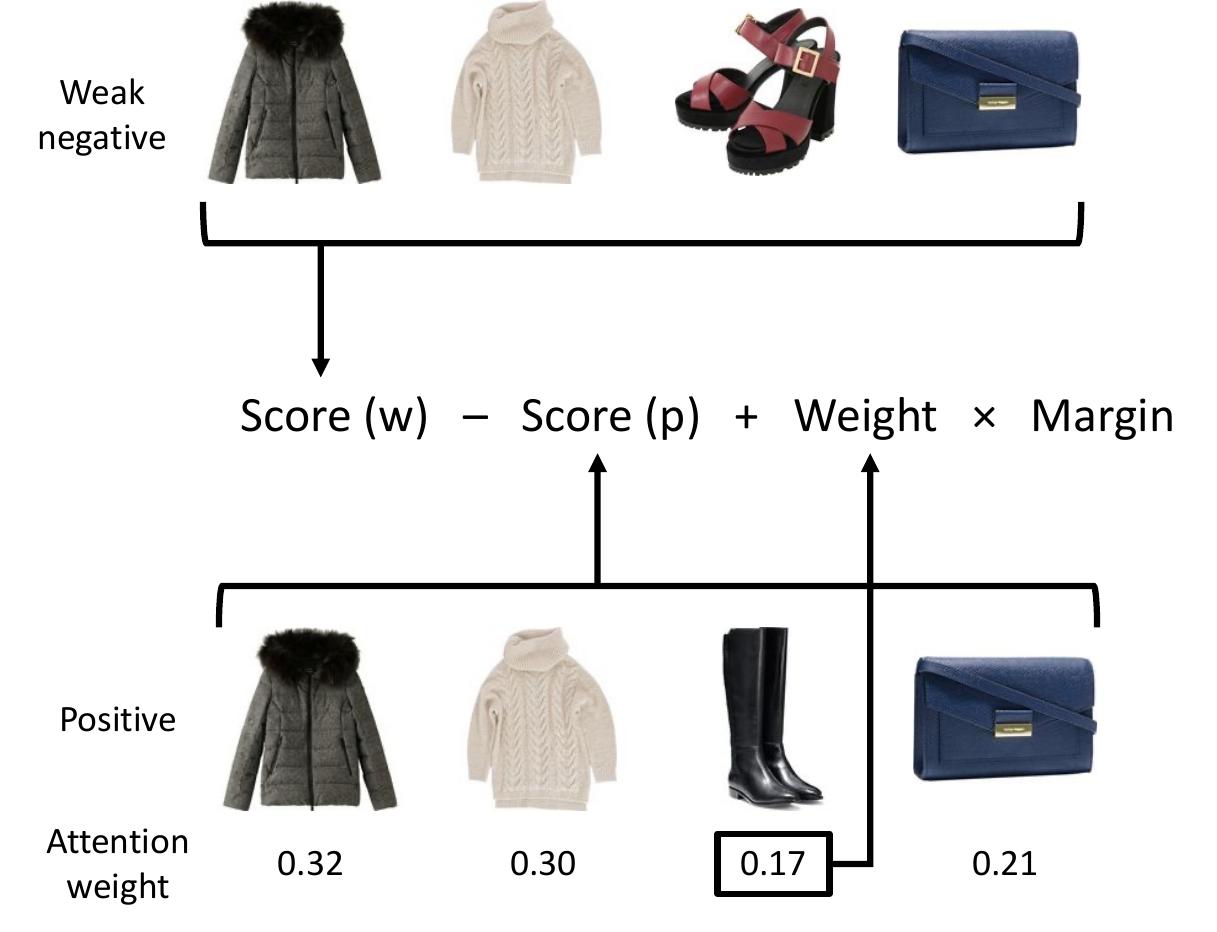}
\caption{Illustration of the adaptive margin loss. Score (w) indicates the compatibility score of a weak negative outfit, and Score (p) indicates the compatibility score of a positive outfit. The margin is weighted by the learned attention weight of the switched item representing its importance in determining the compatibility of an outfit.}
\Description{Adaptive margin loss.}
\label{fig:adaptive margin}
\end{figure}

Although Eq.~\eqref{eq:margin loss} allows the model to learn the compatibility score of the weak negative sample, the margin is fixed for all weak negative samples. This neglects each items contribution to the overall compatibility of an outfit. The margin is fixed whether a jacket among three items or an earring among ten items is switched. Intuitively, the margin should be larger if a more important item is switched and smaller if a less important item is switched. We incorporate
this intuition into the loss by using the learned attention weight in Eq.~\eqref{eq:outfit embedding} to represents the importance of each item in computing the compatibility score at no additional computation cost. We define the \textbf{adaptive margin loss} as:
\begin{equation}
    \mathcal{L}_{AM}\left(B\right) = \sum_{k \in B} (0,-p_p^k+p_w^k+m \cdot A^p_i)
    \label{eq:L_AM}
\end{equation}
where $i$ is the index of the switched item in the weak negative sample and $A^p_i$ is that item's attention weight. We demonstrate this in Fig.~\ref{fig:adaptive margin}.

The overall training loss is defined as follows:
\begin{equation}
    \mathcal{L} = \sum_{B}\left[ c_{FL} \cdot \mathcal{L}_{FL}\left(B\right) + c_{CL} \cdot \mathcal{L}_{CL}\left(B\right) + c_{AM} \cdot \mathcal{L}_{AM}\left(B\right)\right]
\end{equation}
where $c_{CL}$ and $c_{AM}$ are balancing coefficients.

\section{Experiments}
Our experiments aim to address the following questions:
\begin{itemize}
    \item \textbf{Q1}: How does our proposed method perform on personalised outfit recommendation compared to baselines?
    \item \textbf{Q2}: How does contrastive learning improve the model's performance on the outfit recommendation?
    \item \textbf{Q3}: How does the number of history outfits affect the model's performance on the outfit recommendation?
    \item \textbf{Q4}: How does our proposed method perform on item recommendation given a partial outfit?
    \item \textbf{Q5}: How effective is the adaptive margin loss on the item recommendation?
\end{itemize}


\subsection{Experimental Settings}

\subsubsection{Datasets}
We used two popular fashion datasets, IQON3000~\cite{song2019gp} and Polyvore-630~\cite{Lu2019Learning}, for the evaluation. In both datasets, outfits were created by shoppers, which clearly reflects each shopper's preferred outfit styles. We consider outfits created by the same shopper as the shopper's purchase history. The number of items per outfit varies in IQON3000, while Polyvore-630 contains exactly three items  per outfit. A summary of the datasets is shown in Table~\ref{tab:datasets}. 

\subsubsection{Compared Methods}
In both outfit and item recommendation tasks, we compared our model with five baselines: 
\begin{enumerate}
    \item Bi-LSTM~\cite{han2017learning}: Bi-LSTM for fashion compatibility is a model which learns compatibility of an outfit by sequentially predicting each item conditioned on the previous items.
    \item FHN~\cite{Lu2019Learning}: Fashion hashing network (FHN) is a composite model of hashing modules which learn binary codes of shoppers and items to score item-item compatibility and item-shopper compatibility.
    \item LPAE~\cite{Lu2021personalized}: Learnable personalised anchor embedding (LPAE) is a model which encodes anchors specific to each shopper and general anchors. Uses only images of items.
    \item LPAE-T~\cite{Lu2021personalized}: LPAE-T is a variant of LPAE which leverages both images and title of items by concatenating image features and text features to form item features.
    \item OutfitTransformer~\cite{sarkar2023outfittransformer}: OutfitTransformer is the SOTA outfit recommendation model which leverages transformer encoders to learn the internal compatibility of an outfit. It is not personalised.
\end{enumerate}
As baselines, we selected all the competitive models with published code. All reported numbers are obtained on the original test sets. To assess the significance of observed performance differences between models, we used sampled 10,000 bootstrapped test sets and used them to compute confidence
intervals for HAT's metrics. We consider the difference between a baseline and HAT significant if the baseline's metric lies outside HAT's 95\% confidence interval. Confidence intervals (CIs) are shown in Tables~\ref{tab:CP performance} and~\ref{tab:FITB performance}.

\subsubsection{Implementation Details}
For all the baselines and our model, we used frozen CLIP~\cite{radford2021learning} ViT-B/32 weights ($\theta, \phi$) for the image encoder and the text encoder without fine-tuning. We stacked a learnable MLP on top of the CLIP
embeddings to adjust the embeddings without having to fine-tune the larger CLIP models. We set the maximum number of history outfits as ten for all the experiments and the hyperparameters as $\alpha=0.5$, $\gamma=2$, $\tau=1$, $m=5$, $c_{CL}=1$, and $c_{AM}=0.5$. We used AdamW~\cite{loshchilov2018decoupled} optimiser and all experiments were conducted in PyTorch framework~\cite{paszke2019pytorch}.

\begin{table}[t]
  \caption{The number of shoppers, outfits, and fashion items in IQON3000 and Polyvore-630.}
  \label{tab:datasets}
  \begin{tabular}{cccc}
    \toprule
    Dataset & \# Shoppers & \# Outfits & \# Items\\
    \midrule
    IQON3000~\cite{song2019gp} &  3,568 &  308,747 & 672,335 \\
    Polyvore-630~\cite{Lu2019Learning} &  630 & 150,380 & 205,234 \\
  \bottomrule
\end{tabular}
\end{table}

\begin{table*}[!t]
  \caption{AUC comparison of CP with IQON3000 and Polyvore-630.}
  \label{tab:CP performance}
  \begin{tabular}{ccccc}
    \toprule
    {} & \multicolumn{2}{c}{IQON3000} & \multicolumn{2}{c}{Polyvore-630}\\
    \cmidrule(l){2-3} \cmidrule(l){4-5}
    Model & CP-Random (AUC) & CP-Hard (AUC) & CP-Random (AUC) & CP-Hard (AUC) \\
    \midrule
    Bi-LSTM~\cite{han2017learning}  & 0.8909 & 0.4915 & 0.8051 &  0.5158 \\
    FHN~\cite{Lu2019Learning} &  0.9047 & 0.5675 & 0.7490 & 0.4815 \\
    LPAE~\cite{Lu2021personalized} & 0.9510 & 0.5605 & 0.7977 & 0.5127 \\
    LPAE-T~\cite{Lu2021personalized} & 0.9628 & 0.5449 & 0.7872 & 0.4881 \\
    OutfitTransformer~\cite{sarkar2023outfittransformer} & 0.9468 & 0.4650 & 0.7021 & 0.4528 \\
    HAT (ours) & \textbf{0.9665} & \textbf{0.6565} & \textbf{0.8569} & \textbf{0.6159}\\
    HAT 95\% CI & (0.9653, 0.9671)  & (0.6521, 0.6585) & (0.8534, 0.8599) & (0.6109, 0.62310) \\
    \midrule 
    Gain &+0.4\%&+15.7\%&+6.4\%&+19.4\%
    \\
    
  \bottomrule
\end{tabular}
\end{table*}

\subsection{Personalised Outfit Recommendation}
We evaluated the models' performance on personalised outfit recommendation with the \textbf{compatibility prediction (CP)} task. CP evaluates whether a model can determine compatibility of an outfit. It measures the area under the receiver operating characteristic curve (AUC) of compatibility scores given positive samples ($y^o=1$) and negative samples ($y^o=0$).

\subsubsection{Negative Sampling Methods}
For every positive outfit in both datasets, we created negative outfits in two different ways. Firstly, similar to the previous studies on outfit recommendation \cite{sarkar2022outfittransformer}, we randomly switched each item in an outfit to another item within the same item category, resulting in the ratio of positive and negative outfits as 1:1. We call this dataset \textbf{CP-Random}. 

\begin{figure}[t]
\centering
\includegraphics[width=0.43\textwidth]{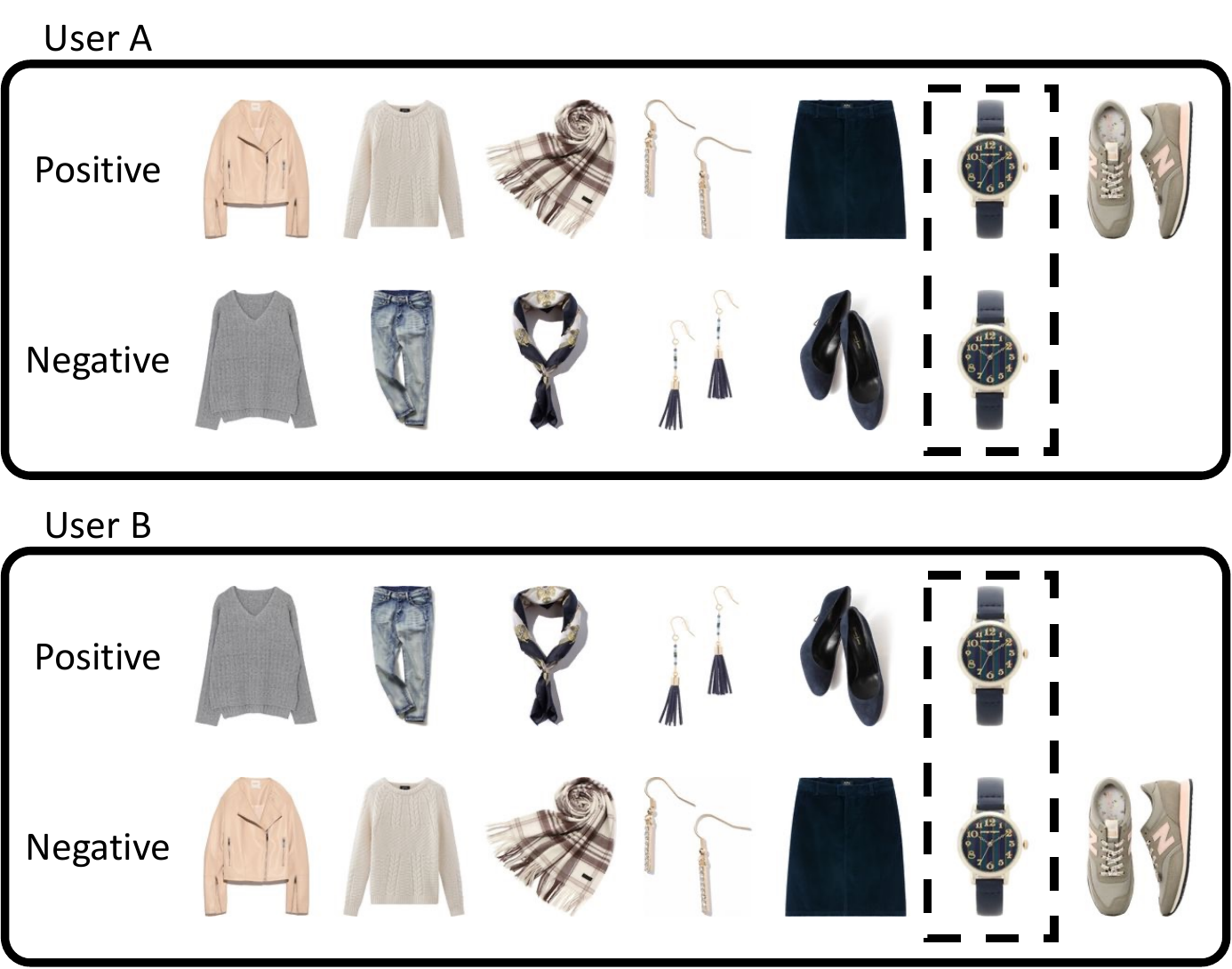}
\caption{An example of a pair of positive outfit and negative outfit in CP-Hard. Shopper A and Shopper B have different positive outfits that share the same watch. For Shopper A, we consider the positive outfit of Shopper B as a negative outfit, and for Shopper B, we consider the positive outfit of Shopper A as a negative outfit.}
\label{fig:cp-hard}
\end{figure}

We reiterate our argument from \ref{sec:adaptive margin loss} that while CP-Random is useful to evaluate the model's ability of distinguishing a random outfit from a compatible outfit, the task is rather easy. In particular,  it is difficult to evaluate a model's personalisation capability since compatible
outfits can be incompatible with a shopper's individual taste, something that CP-Random does not reflect. A personalised recommendation model should avoid recommending outfits that do not align with the shopper's personal style even if the outfits are internally compatible. 

In order to evaluate compatibility of an outfit with a shopper's taste, we treated positive outfits from one shopper as negative outfits for all other shoppers. To avoid cases where a shopper did not purchase outfits because he had not seen them, we only consider those positive outfits from other shoppers that have at least one overlapping item with each outfit. Intuitively, this indicates the two shoppers were exposed to the same  item but purchased other outfits, which implies different fashion styles. We call this dataset \textbf{CP-Hard}. An example is shown in Fig.~\ref{fig:cp-hard}. Note that we trained models only with CP-Random and evaluated them on both CP-Random and CP-Hard. HAT was also trained on
weak negative samples describes in Section~\ref{sec:adaptive margin loss}.

\begin{table}[t]
  \caption{Ablation experiment of contrastive learning.}
  \label{tab:CL ablation}
  \resizebox{0.47\textwidth}{!}{\begin{tabular}{cccc}
    \toprule
    Dataset & Variant & CP-Random (AUC) & CP-Hard (AUC) \\
    \midrule
    \multirow{2}{*}{IQON3000} & Without $\mathcal{L}_{CL}$ & 0.9280 & 0.5350 \\
    & With $\mathcal{L}_{CL}$ & \textbf{0.9665} & \textbf{0.6565} \\
    \midrule
    \multirow{2}{*}{Polyvore-630} & Without $\mathcal{L}_{CL}$ & 0.7183 & 0.4888 \\
    & With $\mathcal{L}_{CL}$ & \textbf{0.8569} & \textbf{0.6159} \\
  \bottomrule
\end{tabular}}
\end{table}

\begin{table*}[!t]
  \caption{Impact of the maximum number of history outfits on personalised outfit recommendation.}
  \label{tab:num history ablation}
  \begin{tabular}{ccccc}
    \toprule
    {} & \multicolumn{2}{c}{IQON3000} & \multicolumn{2}{c}{Polyvore-630}\\
    \cmidrule(l){2-3} \cmidrule(l){4-5}
    \# History outfits & CP-Random (AUC) & CP-Hard (AUC) & CP-Random (AUC) & CP-Hard (AUC) \\
    \midrule
    0 & 0.9511 & 0.4859 & 0.6772 &  0.4652 \\
    10 & \textbf{0.9665} & 0.6565 & \textbf{0.8569} & 0.6159 \\
    20 & 0.9652 & 0.6653 & 0.8485 & 0.6246 \\
    30 & 0.9647 & \textbf{0.6701} & 0.8456 & \textbf{0.6249} \\
  \bottomrule
\end{tabular}
\end{table*}

\begin{table*}[!t]
  \caption{Accuracy comparison of FITB with IQON3000 and Polyvore-630.}
  \label{tab:FITB performance}
  \begin{tabular}{ccccc}
    \toprule
    {} & \multicolumn{2}{c}{IQON3000} & \multicolumn{2}{c}{Polyvore-630}\\
    \cmidrule(l){2-3} \cmidrule(l){4-5}
    Model & FITB-Random (Acc)& FITB-Hard (Acc) & FITB-Random (Acc)& FITB-Hard (Acc)\\
    \midrule
    Bi-LSTM~\cite{han2017learning}  & 0.7258 & 0.5627 & 0.4028 & 0.4554 \\
    FHN~\cite{Lu2019Learning} & 0.5887 & 0.5645 & 0.4252 & 0.4813 \\
    LPAE~\cite{Lu2021personalized} & 0.6739 & 0.5768 & 0.5164 & 0.4887 \\
    LPAE-T~\cite{Lu2021personalized} & 0.6960 & 0.6006 & 0.4889 & 0.4652  \\
    OutfitTransformer~\cite{sarkar2023outfittransformer} & 0.7195 & 0.5969 & 0.3989 & 0.3870 \\
    HAT (ours) & \textbf{0.7406} & \textbf{0.6394} & \textbf{0.5356} & \textbf{0.5363}\\
    HAT 95\% CI & (0.7370, 0.7443) & (0.6355, 0.6434) & (0.5317, 0.5448) & (0.5308,  0.5436) \\
        \midrule
    Gain&+2.0\%&+6.5\%&+3.7\%&+9.7\%\\
  \bottomrule
\end{tabular}
\end{table*}

\subsubsection{Experimental Results (Q1)}
\label{sec:CP results}
Table~\ref{tab:CP performance} shows the experimental results of personalised outfit recommendation. For IQON3000, our method significantly outperforms all baselines. The biggest performance gap between HAT and the second best model is observed in the CP-Hard task where HAT outperforms FHN by 8.9\%. While LPAE-T is the second best model with performance difference of 0.37\% from HAT in CP-Random, the performance difference is much larger in CP-Hard, which is 11.16\%. This highlights HAT's personalisation capability by leveraging shopper purchase history. It also shows that CP-Random is 
inadequate for evaluating personalised outfit recommendation.

A similar trend is observed in Polyvore-630. Our method outperforms the others in every metric, and the biggest performance gap between HAT and the second best model is seen in CP-Hard where HAT outperforms Bi-LSTM by 10.01\%. Despite Bi-LSTM being the simplest baseline, it outperforms more complex models such as LPAE and FHN in CP-Random and CP-Hard. This difference might be due to the smaller dataset size and the simpler outfit configuration where every outfit in Polyvore-630 has exactly three items from three fixed categories while the number of items in an outfit in IQON3000 ranges from 2 to 20 from 59 different categories.

\subsubsection{Effect of Contrastive Learning (Q2)}
\label{sec:Contrastive learning}
We conducted an ablation study on the contrastive loss where the personalised outfit recommendation performance with and without $\mathcal{L}_{CL}$ is shown in Table~\ref{tab:CL ablation}. The performance of both CP-Random and CP-Hard with $\mathcal{L}_{CL}$ is higher than without $\mathcal{L}_{CL}$ for both datasets. This illustrates that contrastive learning is essential in leveraging history outfits properly by aligning outfit representations from the same shopper's history.



\begin{table}[t]
  \caption{Ablation experiments of the adaptive margin loss.}
  \label{tab:adaptive margin ablation}
  \resizebox{0.47\textwidth}{!}{\begin{tabular}{cccc}
    \toprule
    Dataset & Variant & FITB-Random (Acc) & FITB-Hard (Acc) \\
    \midrule
    \multirow{3}{*}{IQON3000} & Without $\mathcal{L}_{AM}$ & 0.6789 & 0.5822 \\
    & Without $A^p_i$ & 0.7257 & 0.6164 \\
    & With $\mathcal{L}_{AM}$, $A^p_i$ & \textbf{0.7406} & \textbf{0.6394} \\
    \midrule
    \multirow{3}{*}{Polyvore-630} & Without $\mathcal{L}_{AM}$ & 0.5139 & 0.5060 \\
    & Without $A^p_i$ & 0.5210 & 0.5160 \\
    & With $\mathcal{L}_{AM}$, $A^p_i$  & \textbf{0.5356} & \textbf{0.5363} \\
  \bottomrule
\end{tabular}}
\end{table}

\subsubsection{Impact of Number of History Outfits (Q3)}
\label{sec:number of history}
The experimental results in Section~\ref{sec:CP results} and Section~\ref{sec:Contrastive learning} clearly demonstrate HAT's personalisation capability when leveraging purchase history. We conducted ablation studies on how the number of history outfits affects the personalisation performance. Table~\ref{tab:num history ablation} shows CP-Random and CP-Hard results with different numbers of history outfits. When the number of history outfits equals to zero, the model is not personalised. In CP-Hard, a higher number of history outfits correlates with better performance on both datasets, indicating the importance of using
all of the purchase history for personalisation. The performance in CP-Random is initially higher with history outfits but decreases as the number of outfits exceeds ten. This shows that while incorporating more history outfits is critical in differentiating different shoppers' personal styles, it may not be as beneficial when distinguishing random outfits.

\subsection{Fill-in-the-blank Task}
The CP task checks how well a model can estimate human compatibility
judgements. For real e-commerce applications, recommending the correct item out of a candidate pool is more important. We therefore evaluate the models' performance on the \textbf{fill-in-the-blank (FITB)} task. FITB evaluates whether a model can complete an outfit where one item is masked and four different choices are given (i.e. one correct item and three wrong items). It measures the accuracy of the model in selecting the correct item for the missing spot. We did not further train the models for this task. Instead, we estimated the CP scores of the four possible outfits, and if the score of the ground-truth outfit is highest, it was considered as a correct prediction.

Notice that our version of FITB requires the model to pick the correct item
without any additional information. Previous work \cite{sarkar2023outfittransformer} has provided the title of the target
outfit as input, making FITB an image retrieval task. Having access to the
target title is not a realistic scenario in e-commerce applications and we
thus do not provide the title in our evaluations. This explains why the
FITB results we report for the Outfit Transformer on Polyvore are lower
than those reported in \cite{sarkar2023outfittransformer}.

\subsubsection{Negative Sampling Methods}
We used two methods to sample negative items for the masked position of an outfit. The first method is to randomly select an item regardless of the item category, and the second method is to randomly select an item in the same item category of the target item. We call the first dataset  \textbf{FITB-Random} and the second one \textbf{FITB-Hard}, as we consider it a more difficult task.

\subsubsection{Experimental Results (Q4)}
Table~\ref{tab:FITB performance} shows test accuracy of FITB. In both datasets, HAT significantly outperforms the other models. Unsurprisingly, FITB-Random accuracy is higher than FITB-Hard accuracy for every model in IQON3000. However, in Polyvore-630, half of the models performed better in FITB-Random, and the other half performed better in FITB-Hard. This result might be caused by the limited number (3) of item categories in Polyvore-630, which results in the many of negative items in FITB-Random having the same item category with the missing item. On expectation, 33\% of FITB-Random outfits are in fact FITB-Hard outfits in Polyvore. Accordingly, the difference in difficulty between FITB-Random and FITB-Hard in Polyvore-630 is less evident than IQON3000.

\subsubsection{Ablation Studies on Adaptive Margin Loss (Q5)}
We conducted two ablation studies on the adaptive margin loss. The first study is to test the effectiveness of introducing weak negative samples where we trained the model with and without $\mathcal{L}_{AM}$. The second ablation study is to examine whether using the attention weight $A^p_i$ improves the performance compared to the non-adaptive margin loss (i.e., setting $A^p_i=1$ in Eq.~\eqref{eq:L_AM}). Table~\ref{tab:adaptive margin ablation} shows that the model without $\mathcal{L}_{AM}$ performs the worst and the model with $\mathcal{L}_{AM}$ and $A^p_i$ performs the best in both datasets. This highlights that having weak negatives in combination with an adaptive margin loss is crucial in FITB. This finding corroborates our intuition that weak negatives provide
a stronger learning signal to the model since they are harder to 
distinguish from positive outfits.

\section{Conclusion}

In this work, we have introduced the History-Aware Transformer (HAT) model for personalised
outfit compatibility prediction. Making compatibility models history-aware
is a straightforward way of personalising outfit recommendations. 
We believe that this is highly relevant for practical use cases since online
shoppers prefer personalised recommendations over generic ones. 

We have also shown how to improve outfit recommendation models by
introducing weak negative samples in the training process, i.e. outfits
where we switch one item at random. This makes weak negatives harder to
distinguish from outfits that are known to be compatible, thus providing
a stronger learning signal to the model. The key to incorporating these
weak negatives into model training is an adaptive margin loss whose 
margin increases with the attention weight for the switched item.

Our experiments show that the proposed model outperforms existing
baselines in scoring outfits as well as in choosing a missing item to
complete an outfit. In several ablation studies, we demonstrate the 
necessity of the weak negatives and the margin loss to the personalisation performance of HAT. The ablations also highlight the importance of aligning outfit representations via supervised contrastive training. That being said, there are several shortcomings to the proposed
model that leave ample room for future research.

First, our model does not actually assemble outfits but only scores or
completes them. However, the model does produce outfit
representations that can be used in retrieval tasks. Future work will focus
on using these representations to retrieve relevant complementary items for 
a base item and build outfits that way. Doing this at scale and choosing
from potentially millions of items is a challenge in itself.

Second, the proposed model does not take the sequence of the customer
history into account. A purchase made a year ago is arguably less
informative of a shopper's current preferences than their most recent 
purchase. Due to the contrastive loss we employ, our model treats all
outfits in the history equally. A common way to account for the diminishing
informativeness of older items is to introduce an exponentially decaying
weighting scheme. We could use such a scheme to downweight
the importance of older outfits in a shopper's history. This is an area
of future investigation.

Finally, the way we represent fashion items is currently restricted to
titles and images. Many e-commerce marketplaces offer much more information
to shoppers. There are customer reviews, star ratings, size and material
information etc. We believe that incorporating this kind of data will
help our model make more accurate compatibility judgements. 

\bibliographystyle{ACM-Reference-Format}
\bibliography{main}


\end{document}